%% file: sigir2019-ltr-tutorial-proposal.tex
\documentclass[sigconf,screen=true,natbib,9pt]{acmart}

\renewcommand\footnotetextcopyrightpermission[1]{} % removes footnote with conference information in first column
\usepackage{acronym}
\usepackage{amsmath}
\usepackage{amssymb}
\usepackage{booktabs}
\usepackage{graphicx}
\usepackage{url}
\usepackage{times}
\usepackage{microtype}
\usepackage{tabularx}
\usepackage{color}
\usepackage[normalem]{ulem}
\usepackage[english]{babel}
\usepackage[misc,geometry]{ifsym} 
\usepackage{enumitem}

\acrodef{IR}{Information Retrieval}
\acrodef{LTR}{Learning to Rank}
\acrodef{OLTR}{Online Learning to Rank}
\acrodef{CLTR}{Counterfactual Learning to Rank}
\acrodef{DBGD}{Dueling Bandit Gradient Descent}
\acrodef{MGD}{Multileave Gradient Descent}
\acrodef{RL}{Reinforcement Learning}
\acrodef{PDGD}{Pairwise Differentiable Gradient Descent}
\acrodef{IPS}{Inverse Propensity Scoring}

\newfont{\mycrnotice}{ptmr8t at 7pt}
\newfont{\myconfname}{ptmri8t at 7pt}

\begin{document}

%\title{Learning to Rank from User Interactions: Counterfactual, Online, and Reinforcement Learning Approaches}
\title{Unbiased Learning to Rank:\\ Counterfactual and Online Approaches}
\subtitle{Tutorial Overview}

\author{Harrie Oosterhuis}
\affiliation{%
\institution{University of Amsterdam}
\city{Amsterdam}
\country{The Netherlands}
}
\email{oosterhuis@uva.nl}

\author{Rolf Jagerman}
\affiliation{%
\institution{University of Amsterdam}
\city{Amsterdam}
\country{The Netherlands}
}
\email{rolf.jagerman@uva.nl}

\author{Maarten de Rijke}
\affiliation{%
\institution{University of Amsterdam}
\city{Amsterdam}
\country{The Netherlands}
}
\email{derijke@uva.nl}

\begin{abstract}
This tutorial covers and contrasts the two main methodologies in unbiased \ac{LTR}: Counterfactual~\ac{LTR} and Online~\ac{LTR}.
There has long been an interest in \ac{LTR} from user interactions, however, this form of implicit feedback is very biased.
In recent years, unbiased \ac{LTR} methods have been introduced to remove the effect of different types of bias caused by user-behavior in search.
For instance, a well addressed type of bias is position bias: the rank at which a document is displayed heavily affects the interactions it receives.
Counterfactual \ac{LTR} methods deal with such types of bias by learning from historical interactions while correcting for the effect of the explicitly modelled biases.
Online \ac{LTR} does not use an explicit user model, in contrast, it learns through an interactive process where randomized results are displayed to the user.
Through randomization the effect of different types of bias can be removed from the learning process.
Though both methodologies lead to unbiased \ac{LTR}, their approaches differ considerably, furthermore, so do their theoretical guarantees, empirical results, effects on the user experience during learning, and applicability.
Consequently, for practitioners the choice between the two is very substantial.
By providing an overview of both approaches and contrasting them, we aim to provide an essential guide to unbiased \ac{LTR} so as to aid in understanding and choosing between methodologies.
\end{abstract}

\settopmatter{printacmref=false}

\maketitle

% reset the acronyms (correct for abstract)
\acresetall

%\section*{Cover Sheet}
%\section{Talk length}
%The tutorial will take a half-day: three hours plus breaks.
%
%\section{Intended Audience}
%Introductory: the audience should be familiar with supervised learning to rank already.
%Previous knowledge about learning from user interactions is not required.
%
%\input{sections/speakers}
%
%\clearpage
%\setcounter{section}{0}

\input{sections/motivation}
\input{sections/objectives}
\input{sections/relevance}
\input{sections/schedule}

\section{Supplied Material}
The slides will be made available to the public,\footnote{SIGIR'19 slides will be published on: \url{http://ltr-tutorial-sigir19.isti.cnr.it/}} we will include references to open source code from related work.

%\section{Relevant references to support proposal evaluation}
%
%\textbf{References on the importance of online evaluation, online learning to rank}\\
%In the last nine years a large body of research shows there is a lot of interest in these fields
%\cite{schuth2016multileave, schuth2015probabilistic, oosterhuis2016probabilistic, oosterhuis2017balancing, oosterhuis2017sensitive, li2018optimizing, jagermanery, schuth2014multileaved, hofmann2013fidelity, radlinski2013optimized, sanderson2010test, hofmann2016online, kohavi2009controlled, kohavi2013online, radlinski2008does, joachims2002optimizing, chapelle2012large, schuth2015predicting, liu2009learning, yue2009interactively, yue2010beyond, hofmann2013balancing, hofmann2013reusing, schuth2014optimizing}.
%Moreover, interest from the industry is has also tremendously increased, with applications from companies such as Bloomberg, Google, Yandex, and Microsoft.

\begin{acks}
This research was partially supported by
Ahold Delhaize,
the Association of Universities in the Netherlands (VSNU),
the Innovation Center for Artificial Intelligence (ICAI),
the Netherlands Organisation for Scientific Research (NWO)
under pro\-ject nr
612.\-001.\-551.
All content represents the opinion of the authors, which is not necessarily shared or endorsed by their respective employers and/or sponsors.

\end{acks}

\bibliographystyle{ACM-Reference-Format}
\bibliography{sigir2019-ltr-tutorial-proposal}

\end{document}

%% file: sections/motivation.tex
% !TEX root = ../sigir2019-ltr-tutorial-proposal.tex

\section{Introduction}

\ac{LTR} has long been a core task in \ac{IR}, as ranking models form the basis of most search and recommendation systems.
Traditionally, \ac{LTR} has been approached as a supervised task where there is a dataset with perfect relevance annotations~\citep{liu2009learning}.
However, over time the limitations of this approach have become apparent. Most importantly, datasets are very expensive to create~\citep{chapelle2011yahoo} and user preferences do not necessarily align with the annotations~\cite{sanderson2010test}.
As a result, interest in \ac{LTR} from user interactions has increased significantly in recent years.

User interactions, often in the form of user clicks, provide \emph{implicit feedback}~\citep{joachims2002optimizing}, and while cheap to collect, they are also heavily biased~\citep{yue2010beyond, wang2018position}.
The most prominent form of bias in ranking is \emph{position bias}: users spend more attention to higher ranked documents, and consequently, the order in which documents are displayed considerably affects the interactions that take place~\citep{wang2018position}.
Another common form of bias is \emph{item selection bias}: users can only interact with documents that are displayed, and as a result, the selection of displayed documents heavily affects which interactions are possible.
Naively ignoring these biases during the learning process will result in biased ranking models that are not optimal for user preferences~\citep{joachims2017unbiased}.
Thus, the field of \ac{LTR} from user interactions is mainly focussed on methods that remove biases from the learning process, resulting in unbiased \ac{LTR}.

The first approach to unbiased \ac{LTR} is \ac{CLTR}; it has its roots in user modeling~\cite{chuklin2015click}.
\ac{CLTR} relies on a user model that models observance probabilities explicitly; this model can be inferred separately~\citep{joachims2017unbiased, agarwal2018consistent, carterette2018offline} or jointly learned~\citep{wang2016learning,ai2018unbiased}.
By adjusting for observance probabilities, the effect of position bias can be removed from learning.
This approach allows unbiased learning from historical data, i.e., interactions collected in the past, as long as an accurate user model can be inferred.

The second approach is \ac{OLTR}, which optimizes by directly interacting with users~\citep{yue2009interactively}.
Repeatedly, an \ac{OLTR} method presents a user with a ranking, observes their interactions, and updates its ranking model accordingly.
Initially, these methods were based around interleaving methods~\cite{joachims2003evaluating} that compare rankers unbiasedly from clicks.
\ac{DBGD} compares its current ranking model with a slight variation at each step, and updates toward the variation if such a preference is inferred~\citep{yue2009interactively}.
This approach is related to existing bandit methods for online learning to re-rank~\cite{katariya2016dcm,kveton2015cascading,lagree2016multiple}.
In contrast with \ac{DBGD}, these reranking approaches do not learn ranking models that can be applied to unseen document and queries.
While \ac{DBGD} has long formed the basis of \ac{OLTR}~\cite{oosterhuis2016probabilistic, oosterhuis2017balancing, schuth2016multileave, hofmann2013reusing, hofmann2013balancing, zhao2016constructing}, recently fundamental problems with this approach were discovered~\cite{oosterhuis2019optimizing}.
As a result, an alternative approach to \ac{OLTR} was proposed: \ac{PDGD}~\citep{oosterhuis2018differentiable}.
By not building on the Dueling Bandit approach \ac{PDGD} avoids the problems recognized with \ac{DBGD} while also displaying considerable performance gains.
%Instead, the \ac{PDGD} method is more akin to policy gradients in \ac{RL}~\cite{sutton1998introduction}.
Thus \ac{OLTR} promises a responsive learning process where ranking systems adapt to users automatically and continuously.

We see that a large shift in unbiased \ac{LTR} has taken place in the last three years: the emergence of \ac{CLTR} from the field of user modelling and the replacement of the \ac{DBGD} approach with \ac{PDGD} in \ac{OLTR}.
It is very important that practitioners and academics have a good understanding of each approach, their advantages, and limitations.
Each approach has different theoretical properties and empirical findings show substantial performance differences depending on the circumstances.
As a result, it is essential for \ac{LTR} practitioners to understand the applicability and effectiveness of each method.
As the field has recently advanced in these different directions, we argue this is the perfect time for a single tutorial to present the two approaches together to the \ac{IR} community. 

In this tutorial, we provide an overview of both \ac{CLTR} and \ac{OLTR} approaches and their underlying theory.
We discuss the situations for which each approach has been designed, and the places were they are applicable.
Furthermore, we compare the properties of the both approaches and give guidance on how the decision between them should be made.
For the field of \ac{IR} we aim to provide an essential guide on unbiased \ac{LTR} to understanding and choosing between methodologies.

%% file: sections/objectives.tex
% !TEX root = ../sigir2019-ltr-tutorial-proposal.tex

\section{Objectives}
The main objectives we wish to achieve with this tutorial are:
\begin{itemize}
\item Motivate the concept of unbiased \ac{LTR}.
\item Provide a complete overview of the two main approaches to unbiased \ac{LTR}.
\item Contrast the theoretical differences between the approaches, show the different fundamental assumptions they make.
\item Give guidance on how a decision between the two approaches should be made, discuss their strengths and weaknesses and what conditions should be considered when deciding between them.
\item Discuss future directions for unbiased \ac{LTR}.
\end{itemize}

%% file: sections/relevance.tex
% !TEX root = ../sigir2019-ltr-tutorial-proposal.tex

\section{Relevance to the IR community}
%Relevance to the information retrieval community and reference to tutorials in the same area at SIGIR or related conferences (including WSDM, WWW, KDD, ACL, etc.). 

Many open questions remain to be addressed and there are many opportunities for the information retrieval community to benefit from and contribute to the area. 
Ever since the first publications on learning to rank (such as, e.g.,~\cite{fuhr1991probabilistic}), the well-known information retrieval conferences, such as SIGIR, CIKM, ECIR, WSDM, WWW, have seen follow-up work, as have related conferences, such as KDD, ICML, and NIPS. We estimate that in the last five years alone, hundreds of papers have been published on learning to rank.

As far as we are aware there has been no tutorial on \emph{unbiased} LTR that brings the two angles (counterfactual and online) together, neither at SIGIR nor at any of the conferences listed above.
There have been tutorials on counterfactual \ac{LTR}, cf.~\cite{ai2018unbiased,joachims2016counterfactual}, but they ignore online \ac{LTR}.
Similarly, existing tutorials on online \ac{LTR}, cf.~\cite{grotov2016online, oosterhuis-2018-online}  mostly ignore counterfactual \ac{LTR}.
Therefore, it appears this is the first tutorial to discuss and contrast both unbiased \ac{LTR} methodologies comprehensively.

%% file: sections/schedule.tex
% !TEX root = ../sigir2019-ltr-tutorial-proposal.tex

\section{Format and Detailed Schedule}
The tutorial will consists of two hours of lectures, split in two one-hour blocks by breaks.

\subsection*{Introduction (10 min)}
Brief introduction on the limitations of supervised learning to rank, and biases in user interactions, so that the audience understands the need for unbiased \ac{LTR}.
\begin{itemize}[label={}]
\item \textbf{5 min -- Limitations of the supervised approach}\\
Discuss the limitations of using annotated datasets~\citep{liu2009learning}, most importantly: they are expensive~\cite{chapelle2011yahoo}, they do not necessarily agree with users~\cite{sanderson2010test}, and in some situations such a dataset cannot be constructed~\cite{wang2016learning}.
\item \textbf{5 min -- Learning from user interactions}\\
User interactions provide an alluring alternative: by learning from their behavior the \emph{true} preferences of users may be found~\citep{radlinski2008does, joachims2002optimizing}.
However, user interactions contain noise and biases~\citep{yue2010beyond}, for reliable \ac{LTR} \emph{position bias} has to be countered.
Similarly, in many places \emph{selection bias} is unavoidable and has to be dealt with.
\end{itemize}

\subsection*{Counterfactual Learning to Rank (50 min)}
The \ac{CLTR} approach uses explicit user models to infer the probability that a document was observed separately.
These observance probabilities then can be used to counter the effect of position bias.
\begin{itemize}[label={}]
\item \textbf{15 min -- Counterfactual evaluation}\\
Discuss the offline evaluation of online metrics using \acf{IPS}.
We present the proof that \ac{IPS} produces an unbiased estimate.
\ac{IPS} is the tool that underlies all of the \ac{CLTR} methods, and it is important for the audience to have a good grasp of it.
\item \textbf{10 min -- Propensity-weighted \ac{LTR}} \\
Describe in detail propensity-weighted \ac{LTR} methods~\cite{joachims2017unbiased,wang2016learning,bendersky2018learning}.
Discuss the assumptions made by these methods and walk through the algorithms step-by-step.
\item \textbf{15 min -- Estimating position bias} \\
Discuss position bias estimation techniques~\cite{wang2018position}, which are necessary to compute the propensity scores used in all \ac{IPS}-based learning algorithms.
We focus on both online estimation of position bias~\cite{wang2018position} and offline estimation of position bias~\cite{agarwal2018consistent}.
Additionally, we briefly look at trust-bias and how it can be addressed~\citep{agarwal2019addressing}.
\item \textbf{10 min -- Practical considerations} \\
Highlight some of the practical difficulties and their solutions, such as high variance~\cite{swaminathan2015counterfactual}.
\end{itemize}

\subsection*{Online Learning to Rank (45 min)}
\ac{OLTR} methods learn by directly interacting with users, they deal with biases by adding stochasticity to the displayed results.
\begin{itemize}[label={}]
\item \textbf{5 min -- Online evaluation} \\
Discuss interleaving and how it deals with position bias~\citep{joachims2003evaluating, hofmann2013fidelity}.
Most of the initial \ac{OLTR} methods rely on interleaving; it is important the audience understands this basis.
\item \textbf{10 min -- Dueling Bandit Gradient Descent}\\
Describe \ac{DBGD}: the original \ac{OLTR} method~\citep{yue2009interactively} which is based on interleaving.
This method defined a decade of \ac{OLTR} algorithms.
\item \textbf{5 min -- Extensions of DBGD and their limitations} \\
Many extensions of \ac{DBGD} have been proposed~\cite{oosterhuis2016probabilistic, oosterhuis2017balancing, schuth2016multileave, hofmann2013reusing, hofmann2013balancing, zhao2016constructing}, we will briefly describe some approaches and show that they do not lead to long-term improvements in performance.
\item \textbf{10 min -- Regret bounds of DBGD and their problems} \\
The regret bounds of \ac{DBGD} guarantee that its performance should eventually approximate the optimal performance. However, empirically we do not observe this behavior~\citep{schuth2016multileave, oosterhuis2018differentiable}. Recent work has found that the regret bounds rely on assumptions which are impossible for ranking problems~\citep{oosterhuis2019optimizing}. Understanding these issues may be very valuable for future work searching for regret bounds for ranking problems.
\item \textbf{10 min -- Pairwise Differentiable Gradient Descent} \\
Latest \ac{OLTR} method that does not rely on \ac{DBGD}. Optimizes a probabilistic policy and deals with bias with some randomization in results. Proved to be unbiased w.r.t. position and selection bias~\citep{oosterhuis2018differentiable}.
\item \textbf{10 min -- Comparison of PDGD and DBGD} \\
Discuss empirical comparisons between \ac{PDGD} and \ac{DBGD} which show \ac{PDGD} outperforming \ac{DBGD} in all experimental conditions~\citep{oosterhuis2018differentiable, oosterhuis2019optimizing}.
Compare \ac{PDGD} and \ac{DBGD} on a theoretical level to explain these differences.
\end{itemize}

\subsection*{Conclusion (15 min)}
Conclude the tutorial by summarizing the previous sections and fully comparing and contrasting the three different approaches.
\begin{itemize}[label={}]
\item \textbf{10 min -- Summarize the two methodologies and their differences} \\
Reflect on the two approaches to unbiased \ac{LTR}, contrast their properties and applicability.
Consider differences in theoretical properties and empirically observed performance~\citep{jagerman2019comparison}.
Recognize in which situations each method is more suited.
\item \textbf{5 min -- Future directions for unbiased learning to rank} \\
We draw a picture of what current \ac{LTR} methods can do for current applications, then, we identify problems with the current approach and speculate what potential solutions may look like.
We finish by describing the promising directions that future \ac{LTR} work could investigate.
\end{itemize}